\documentclass{emulateapj}
\usepackage{lscape}
\usepackage{rotate}
\usepackage{longtable}
\usepackage{natbib}
\shorttitle{X-ray Selected AGN Hosts in the Cl1604 Supercluster}
\shortauthors{Kocevski et al.}

\begin{document}

\title{Properties of Galaxies Hosting X-ray Selected Active Galactic Nuclei in the Cl1604 Supercluster at $z=0.9$}
\author{Dale D. Kocevski, Lori M. Lubin, Brian C. Lemaux, Roy Gal\altaffilmark{1}, Christopher D. Fassnacht, Robin Lin\altaffilmark{2}, Gordon K. Squires\altaffilmark{3}}

\affil{Department of Physics, University of California, Davis, 1 Shields Avenue,  Davis, CA 95616}
\altaffiltext{1}{Institute for Astronomy, University of Hawaii, 2680 Woodlawn Dr., Honolulu, HI 96822}
\altaffiltext{2}{Siemens Medical Solutions USA, Inc., 1230 Shorebird Way, MS 6-1, Mountain View CA 94043-1344}  
\altaffiltext{3}{\emph{Spitzer} Science Center, M/S 220-6, California 
                 Institute of Technology, 1200 East California Blvd, Pasadena, CA 91125}
\email{kocevski@physics.ucdavis.edu}

\begin{abstract}
Recent galaxy evolution models suggest that feedback from Active Galactic
Nuclei (AGN) may be responsible for suppressing star formation in their host
galaxies and the subsequent migration of these systems onto the red sequence.  
To investigate the role of AGN in driving the evolution of their hosts, we
have carried out a study of the environments and optical properties of
galaxies harboring X-ray luminous AGN in the Cl1604 supercluster at
$z\sim0.9$.  Making use of \emph{Chandra}, \emph{HST}/ACS and
\emph{Keck}/DEIMOS observations, we examine the integrated colors,
morphologies and spectral properties of nine moderate-luminosity ($L_{\rm X}
\sim 10^{43}$ erg s$^{-1}$) type 2 Seyferts detected in the Cl1604 complex. 
We find that the AGN are predominantly hosted by luminous spheroids and/or
bulge dominated galaxies which have colors that place them in the
valley between the blue cloud and red sequence in color-magnitude space,
consistent with predictions that AGN hosts should constitute a transition
population.  Half of the hosts have bluer overall colors as a result of blue
resolved cores in otherwise red spheroids and a majority show signs of recent
or pending interactions.  We also find a substantial number exhibit strong
Balmer absorption features indicative of post-starburst galaxies, despite the
fact that we detect narrow [OII] emission lines in all of the host spectra.
If the [OII] lines are due in part to AGN emission, as we suspect, then this
result implies that a significant fraction of these galaxies (44\%) have
experienced an enhanced level of star formation within the last $\sim1$ Gyr which was
rapidly suppressed.  Furthermore we observe that the hosts galaxies tend
to avoid the densest regions of the supercluster and are instead located in
intermediate density environments, such as the infall region of a massive
cluster or in poorer systems undergoing assembly. Overall we find that the
properties of the nine host galaxies are generally consistent with a scenario
in which recent interactions have triggered both increased levels of nuclear
activity and an enhancement of centrally concentrated star formation,
followed by a rapid truncation of the latter, possibly as a result of
feedback from the AGN itself.  Our finding that the hosts of
moderate-luminosity AGN within the Cl1604 supercluster are predominantly a
transition population suggests AGN feedback may play an important role in
accelerating galaxy evolution in large-scale structures.
\end{abstract}


\section{Introduction}

Several studies have demonstrated that galaxies segregate by rest-frame color,
with passively evolving early-type galaxies forming a well defined ``red
sequence'' and gas-rich, star forming systems located in a more diffuse
``blue cloud'' (e.g.~Baldry et al.~2004).   This observed color bimodality is commonly thought to be
the result of an evolutionary sequence, such that galaxies in the blue cloud
quickly migrate onto the red sequence following the termination of star
formation, leading to a sparsely populated transition zone between the two
regions (Faber et al.~2007; although see Driver et al.~2006 for an alternative view).  
While simple passive evolution likely contributes to the migration of
galaxies from blue to red, recent findings that galaxies exhibit reduced star
formation rates as they assemble into regions of even moderate density (Lewis
et al.~2002; Gomez et al.~2003) suggest that processes prevalent in richer
environments must also play a role in terminating star formation and driving
galaxy evolution.  Recently merger-driven feedback from Active Galactic
Nuclei (AGN) has been suggested as one such process (Hopkins et al.~2005,
2007; Somerville et al.~2008). 

In this scenario strong gravitational torques produced as a result of galaxy
mergers funnel material to the nuclear region of a system, leading to
elevated accretion onto the central black hole and enhanced star formation in
the form of a nuclear starburst (Barnes \& Hernquist 1991; Mihos \& Hernquist
1996).  If sufficiently energetic, the AGN eventually drives outflows that
disrupt the host, effectively quenching ongoing star formation by removing
the galaxy's gas supply (Springel et al.~2005; Hopkins et al.~2005).   As the last generation
of newly formed stars fade, the galaxy quickly migrates from the blue cloud
to the red sequence, with the activity of the central black hole gradually
declining over the same period.

This feedback mechanism provides a method by which the overall properties
of a galaxy can be regulated by its central black hole. Indeed there is a
growing body of evidence that suggests AGN are directly linked to the
evolution of their host galaxies.  This includes the observed correlation
between central black hole mass and the stellar velocity dispersion of a host
galaxy's bulge (Gebhardt et al.~2000; Tremaine et al.~2002) and the coeval
decline in both the cosmic star formation rate and nuclear activity since
$z\sim1$ (Boyle \& Terlevich 1998).  

If AGN are directly responsible for driving the evolution of their hosts,
then signs of recent or ongoing transformation should be present in these
galaxies. This includes disturbed morphologies due to recent interactions,
post-starburst signatures resulting from rapidly quenched star formation
following a merger-induced enhancement, and colors consistent with the
sparsely populated transition zone in the color-magnitude diagram.   
While many of these features are observed in the hosts of luminous quasars (QSO),
studies of moderate-luminosity Seyferts ($L_{\rm X}\sim10^{43}$ erg s$^{-1}$) in this regard have produced mixed
results.   For example, while such AGN are predominantly found in massive,
early-type galaxies which have younger stellar populations than their
non-active counterparts (e.g. Kauffmann et al.~2003), only a
small fraction show signs of recent merger activity (De Robertis et al.~1998;
Grogin et al.~2005; Pierce et al.~2007).   Likewise, although AGN are found in post-starburst galaxies
more often than in normal galaxies (Heckman 2004, Yan et al.~2006, Goto et
al.~2006), the fraction of Seyferts associated with post-starburst hosts at low redshift remains fairly
small; on the order of $4\%$ according to a recent study by Goto et al.~(2006). 

A clearer picture of the prevalence of AGN feedback as an evolutionary
driver may be obtained by studying host galaxies in and around large-scale
structures at high redshift.  In addition to a considerably higher AGN density and galaxy
merger rate at $z\sim1$ relative to the present epoch (Barger et al.~2005;
Kartaltepe et al.~2007), there are indications that AGN are triggered more often
in structures than in the field (Gilli et al~2003; Cappelluti et al~2005; Eastman et al.~2007;
Kocevski et al.~2008), presumably due to the presence of richer, dynamic
environments where galaxy interactions are more common (Cavaliere et al.~1992).  
Several recent wide-field surveys have found that the colors of higher-redshift
host galaxies are consistent with a population in transition (Sanchez et
al.~2004; Nandra et al.~2007; Georgakakis et al.~2008; Silverman et al.~2008;
although see also Westoby et al.~2007) and there are
indications the fraction of AGN hosts exhibiting post-starburst signatures
is higher at $z\sim0.8$ (Georgakakis et al.~2008).  Furthermore, Silverman et
al.~(2008) found that a majority of host galaxies associated with two
large-scale structures in the Extended \emph{Chandra} Deep Field South
(E-CDF-S) at $z\sim0.7$ have colors consistent with the transition zone of the
color-magnitude diagram.  Coupled with the increased incidence of AGN in such
regions, this finding implies that AGN related feedback may play an important
role in accelerating galaxy evolution in large-scale structures. 

In this study we examine the environments and optical properties of galaxies
hosting X-ray selected, moderate-luminosity AGN in the Cl1604 supercluster
at $z=0.9$.  The system is the largest structure mapped at redshifts
approaching unity, consisting of eight spectroscopically confirmed galaxy
clusters and groups and a rich network of filamentary structures.  The system
spans roughly 10 $h_{70}^{-1}$ Mpc on the sky and 100 $h_{70}^{-1}$ Mpc in depth. 
The complex structure of the supercluster, as mapped by extensive
spectroscopic observations, is described in Gal et al.~(2008; hereafter G08) and our X-ray
observations of the system are discussed in Kocevski et al.~(2008; hereafter K08).  The
Cl1604 supercluster is well suited for this study as it provides
a diverse set of  structures within which AGN may be preferentially
found and a wide range of environments and local conditions that can help
constrain the mechanisms responsible for triggering their activity.

In the following sections we begin by describing our X-ray and optical observations of
the Cl1604 system (\S2) and the process by which we identify AGN in the
supercluster (\S3).  This is followed by an examination of the morphology
(\S4.1), optical colors (\S4.2), spectral properties (\S4.3), and
environments (\S4.4) of the galaxies found to host AGN.  We also determine
the fraction supercluster members which harbor AGN and compare this to
low-redshift results in \S5.  We discuss our results and
their implications for the AGN feedback model in driving galaxy evolution in
the Cl1604 system in \S6.  Finally we summarize our results in \S7.  Throughout
this paper all quoted line equivalent widths are in the rest frame,
magnitudes are in the AB system and we assume a $\Lambda$CDM cosmology with
$\Omega_{m} = 0.3$, $\Omega_{\Lambda} = 0.7$, and $H_{0} = 70$ $h_{70}$ km s$^{-1}$ Mpc$^{-1}$.

\section{Observations and Data Reduction}
\label{sect-data}

\subsection{X-ray Observations}

 Observations of the Cl1604 supercluster were carried out with {\it
   Chandra's} Advanced CCD Imaging Spectrometer (ACIS; Garmire et al.~2003)
 on 2006 June 23 (obsID 7343), June 25 (obsID 6933), and October 01 (obsID
 6932).  A detailed description of the observations and subsequent data reduction are
 given in K08, therefore we only provide a brief summary
 here.  The three observations of the supercluster consist of two
 pointings, one encompassing the northern portion of the system and the
 other the southern portion, with a $4\farcm9$ overlap between the imaged
 regions.  Each pointing was imaged with the $16\farcm9\times16\farcm9$
 ACIS-I array for total integration times of 19.4, 26.7, and 49.5 ksec for the
 7343, 6933, and 6932 datasets, respectively. All three datasets were
 reprocessed and analyzed using standard CIAO 3.3 software tools and version
 3.2.2 of the {\it Chandra} calibration database available through {\it
   Chandra} X-ray Center (CXC)\footnote{http://cxc.harvard.edu/}.  

We searched for point sources using the wavelet-based {\tt wavdetect}
procedure in CIAO, employing the standard $\sqrt2$$^i$ series of wavelet
pixel scales, with $i=0-16$.  Object detection was carried out separately on
images in each of the 0.5-2 keV (soft), 2-8 keV (hard) and 0.5-8 keV (full)
X-ray bands and the source catalogs cross-correlated, resulting in 265 unique sources detected in the two pointings of
the Cl1604 system.  Source properties, including count
rates and detection significances, were determined with follow-up aperture
photometry on all sources found by {\tt wavdetect}.  The photometry was
carried out on the vignetting-corrected, soft- and hard-band images of both
pointings and an appropriate aperture correction (as determined from the PSF
libraries in the {\it Chandra} calibration database) was applied to the
background-subtracted net counts of each source.  Counts in the full band
were then determined as a sum of the measured net counts in the soft and hard
bands.  Of the 265 sources found by {\tt  wavdetect}, 161 had detection
significances greater than $3\sigma$ in at least one X-ray band.

Source fluxes in the soft and hard bands were determined by normalizing a
power-law spectral model to the net count rate measured for each source.
These rates were determined by dividing the net counts measured in the
vignetting-corrected images by the nominal exposure time at the aimpoint of
each observation.  We assumed a photon index of $\gamma = 1.4$ for the
power-law model and a Galactic neutral hydrogen column density of
$1.21\times10^{20}$ cm$^{-2}$ (Dickey \& Lockman 1990), resulting in a count
rate to unabsorbed-flux conversion factor of $5.85\times10^{-12}$ and
$2.08\times10^{-11}$ erg cm$^{-2}$ s$^{-1}$ per cnts s$^{-1}$ in the
soft and hard bands, respectively.  Full-band fluxes were again determined by
summing the flux in the soft and hard bands.   Rest frame X-ray luminosities
were calculated for sources matched to galaxies with measured redshifts (see
\S\ref{sect-opt-matching}) using the standard luminosity distance equation
and a $(1+z)^{\gamma - 2}$ $k$-correction.  Properties for the 161 unique
point sources detected with at least a $3\sigma$ significance in the field of
Cl1604 are listed in Table 3 of K08.

\subsection{Optical Imaging}

The optical imaging of the supercluster used for this study consists of 17
pointings of the Advanced Camera for Surveys (ACS) on the \emph{Hubble Space
  Telescope (HST)}.  The ACS camera consists of two $2048\times4096$ CCDs
with a pixel scale of 0.05 $^{\prime\prime}$/pixel, resulting in a
$\sim3^{\prime}\times3^{\prime}$ FOV.  The 17 pointing mosaic was designed to
image nine of the ten galaxy density peaks observed in the field of the
supercluster by Gal et al.~(2005, 2008).  Observations were taken in both the
F606W and F814W bands and consist of 15 pointings from GO-11003 (PI Lubin)
with effective exposure times of 1998 sec\footnote{With the exception of one
  pointing which lost guiding due to an incorrect attitude, resulting in a
  useable exposure time of 1505 sec and a gap in the mosaic.}
and 2 GTO pointings from G0-9919 (PI Ford) centered on clusters Cl1604+4304
and Cl1604+4321 with effective exposure time of 4840 sec. Our average
integration time of 1998 sec resulted in completeness depths of $\sim26.5$
mag in each band.  

\begin{figure}[t]
\epsscale{1.1}
\plotone{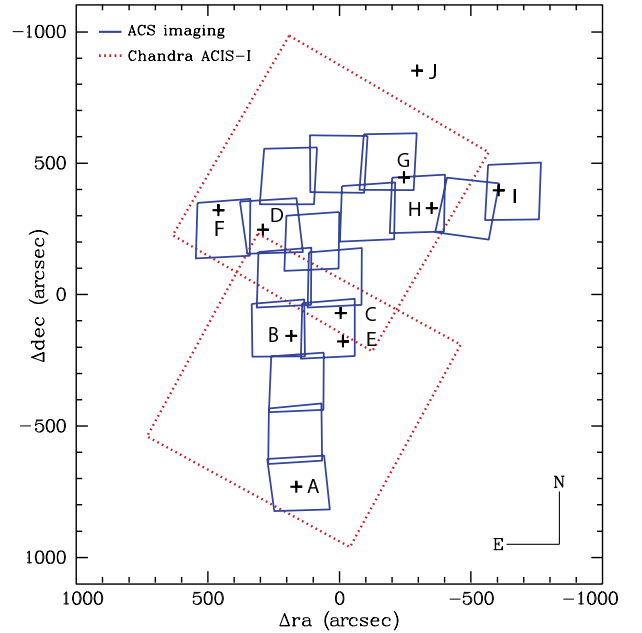}
\caption{Positions and FOVs of the two ACIS-I and 17 ACS pointings
  in the Cl1604 region. The positions of the ten red-galaxy overdensities found by Gal et
  al. (2005, 2008) are marked and labeled following the naming convention of
  Gal et al.~(2005) \label{fig-data_outline}}
\end{figure}

The ACS data were reduced using the pipeline developed by
the HST Archive Galaxy Gravitational Lens Survey (HAGGLES) team.  The
pipeline processing consists of calibrating the raw data using the best
reference files provided by the Hubble Space Telescope archive, subtracting
the background on each chip, iteratively determining the best shifts between
the dithered exposures using multiple calls to {\tt SExtractor} (Bertin \&
Arnouts 1996) and {\tt multidrizzle} (Koekemoer et al.~2002), aligning the
final drizzled image to the USNO-B1 catalog (Monet et al.~2002) and
resampling the images to a pixel scale of 0.03$^{\prime\prime}$/pixel.
Object detection and photometry was carried out using {\tt SExtractor} in dual-image mode,
with the detection image being a weighted average of the F606W and F814W
images.  Detection parameters DETECT\_MINAREA and DETECT\_THRESH were set to
three pixels above $3\sigma$, while the deblending parameters DEBLEND\_NTHRESH
and DEBLEND\_MINCONT were set to 32 and 0.03, respectively.
An outline of the region covered by the ACS mosaic relative to the ACIS-I observations is
shown in Figure \ref{fig-data_outline}. Full details of the HAGGLES pipeline
can be found in Marshall et al.~(2008) and Schrabback et al.~(2008).

\subsection{Optical Spectroscopy}
\label{opt-spect}
The Cl1604 supercluster has been extensively mapped using the
Low-Resolution Imaging Spectrograph (LRIS; Oke et al. 1995) and the
Deep Imaging Multi-object Spectrograph (DEIMOS; Faber et al.~2003) on
the \emph{Keck} 10-m telescopes (Oke, Postman, \& Lubin 1998; Postman, Lubin, \& Oke
1998; Lubin et al.~1998; Gal \& Lubin 2004; G08). The
complex target selection, spectral reduction, and redshift measurements
are described in detail in Section 3 of G08. The final
spectroscopic catalog contains 1,618 unique objects. Redshifts derived for
these objects are given a spectroscopic quality, $Q_{spect}$, between 1 and
4, where 1 indicates that a secure redshift could not be determined due to poor
signal, lack of features or reduction artifacts,  2 is a redshift obtained from either a single
feature or two marginally detected features, 3 is a redshift derived from at
least one secure and one marginal feature, and 4 is assigned to spectra with
redshifts obtained from several high signal-to-noise features.  
$Q_{spect} = -1$ is used for sources securely identified as stars.

In this sample, we find 140 stars and 1,138 extragalactic objects with $Q_{spect} \ge 3$.
A total of 417 galaxies are in the nominal redshift range of
the supercluster between $0.84 \le z \le 0.96$. This extensive
spectroscopic database is larger by a factor of $\sim 10$ than that
for any other known moderate-to-high redshift supercluster.

\section{Identifying Supercluster AGN}
\label{sect-opt-matching}

To find optical counterparts to the 161 X-ray point sources detected
in the field of the Cl1604 supercluster we employed a maximum likelihood
technique described by Sutherland \& Saunders (1992) and more recently
implemented by Taylor (2005) and Gilmour et al.~(2007).  The method gauges
the likelihood that a given optical object is matched to an X-ray source by
comparing the probability of finding a genuine counterpart with the
positional offset and magnitude of the optical candidate relative to that of
finding a similar object by chance.  The advantages of this approach over a
simple nearest neighbor match is that it takes into account the density of
objects as faint as the optical counterpart as well as the distance between
sources and the X-ray positional errors.  The details of the process are
described in K08 and therefore we focus instead on the
results of the matching here.

\begin{figure}[t]
\epsscale{1.05}
\plotone{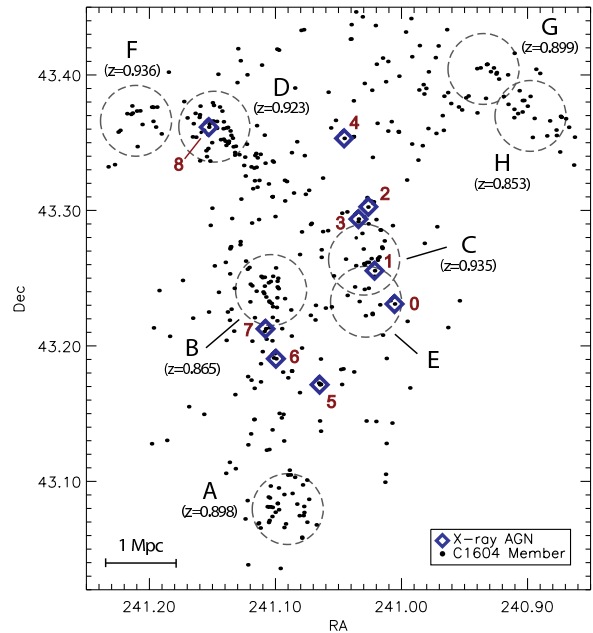}
\caption{Spatial distribution of the 9 X-ray sources that fall within the
  traditional redshift boundaries of the Cl1604 supercluster ($0.84<z<0.96$).  
  Individual clusters and groups in the Cl1604 system are labeled using the
  naming convention of Gal et al.~(2005). The dashed circles extend 0.5
  $h_{70}^{-1}$ Mpc from the center of each cluster. \label{fig-agn-spatial-dist}}
\end{figure}

Using the maximum likelihood technique we have matched 125 optical sources to
our X-ray catalog.  Of these 42 have spectroscopic information available and
we have derived reliable redshifts ($Q_{spect} \ge 3$)
for 35 of the sources.  We find that while the matched sources cover a wide
range of redshifts ($0.055 \le z \le 2.30$), the distribution exhibits a
clear peak at $z=0.9$, the average redshift of the Cl1604 supercluster (see
Figure 6 of K08). A total of nine sources have redshifts
between $0.84<z<0.96$ and fall within the redshift boundaries of
the supercluster.  Eight of these nine
galaxies were imaged by our ACS survey of the supercluster and therefore have
high-resolution imaging available for morphological and resolved color
analyses.  The coordinates and redshifts of the nine supercluster AGN are
listed in Table \ref{tab-prop} and their location in the supercluster are
shown in Figure \ref{fig-agn-spatial-dist}.

Also listed in Table \ref{tab-prop} are the X-ray properties of the nine AGN,
including their rest-frame 0.5-8 keV luminosities and hardness ratios, $HR$,
measured as $HR = \frac{H-S}{H+S}$ where $H$ and $S$ are the net counts in the
hard and soft bands, respectively.  The AGN have luminosities ranging from
0.78 to 4.47 $\times10^{43}$ $h_{70}^{-2}$ erg s$^{-1}$ and have predominantly soft
X-ray spectra; the nine sources have a median $HR$ of $-0.20$.  It should be
noted that the luminosity of all nine AGN exceed the luminosity attributable
to starburst galaxies by over an order of magnitude ($L_{\rm X} \sim
10^{42}$ erg s$^{-1}$; e.g. Bauer et al.~2002), making it highly
unlikely that their X-ray emission is powered exclusively by star formation
activity.   On the other hand, their luminosities fall below the level of
bright QSOs ($L_{\rm X} \sim 10^{44}$ erg s$^{-1}$),
which often dominate the optical emission from their host galaxies.
Silverman et al.~(2005) have shown that the optical emission associated with
AGN with luminosities below $5\times10^{43}$ $h_{70}^{-2}$ erg s$^{-1}$ (0.5-8 keV) is
primarily due to light from the host galaxy.  Since the luminosity of all our AGN fall
below this limit, we proceed under the assumption that the optical properties of
the nine supercluster host galaxies suffer minimal contamination from their active nuclei.

\section{AGN Host Properties}

\subsection{Morphologies}
 
\begin{figure}[t]
\epsscale{1.2}
\plotone{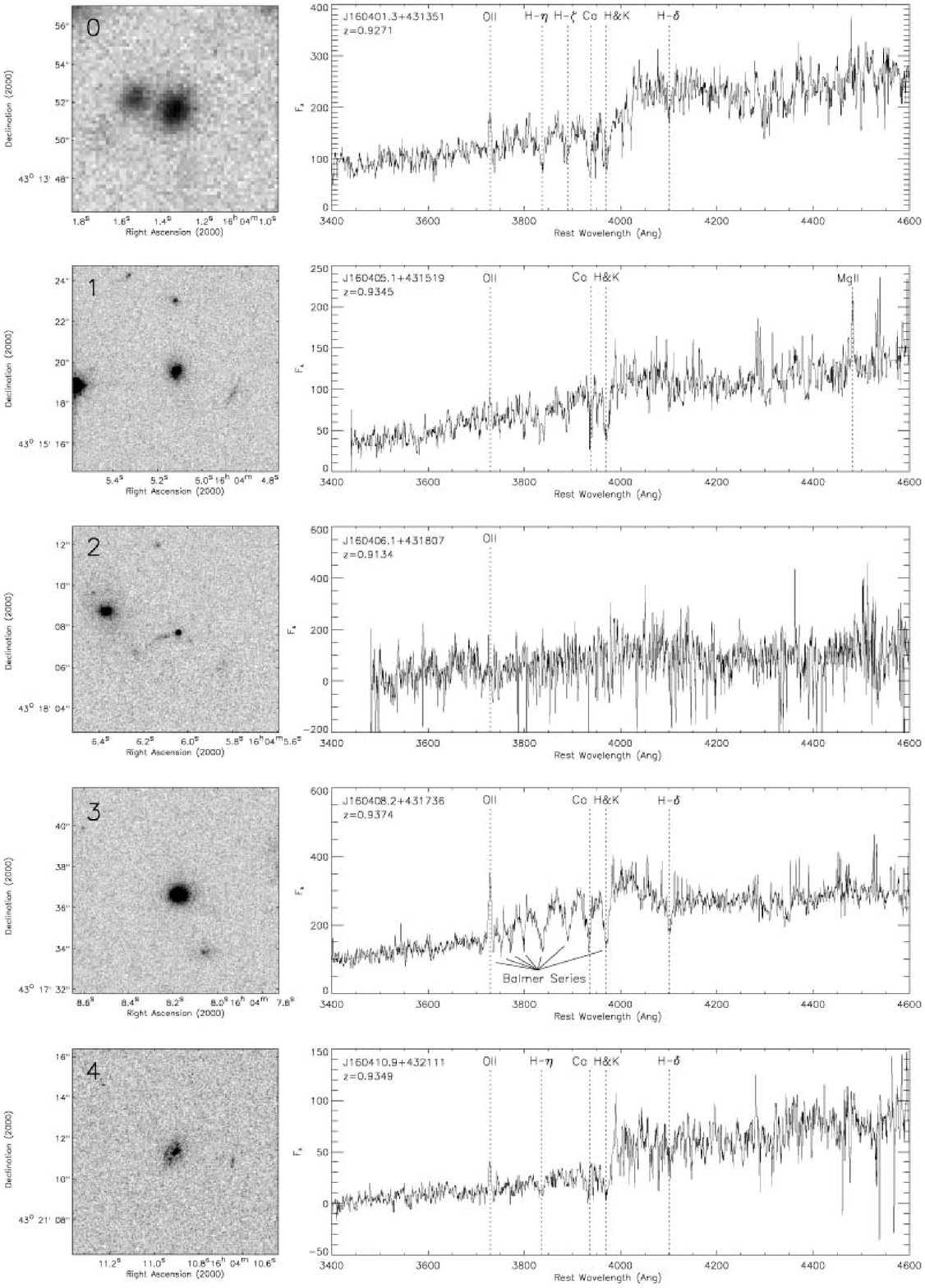}
\caption{(\emph{left}) Thumbnail images of the eight supercluster AGN imaged
         with \emph{HST}/ACS in the F814W band as well as one source imaged with
         \emph{Palomar}/LFC in the $i^{\prime}$ band.  Each image spans roughly
         $10^{\prime\prime}$ on a side and is shown with a log scaling.
         (\emph{right}) Optical spectra of all nine supercluster AGN obtained
         with \emph{Keck}/DEIMOS.  The flux scale is presented in
         self consistent, arbitrary units.  The spectra have typically been
         binned by 10 pixels for clarity, where one pixel equals 0.33
         \AA. \label{fig-thumbs+spect1}}
\end{figure}

\begin{figure}[t]
\epsscale{1.2}
\plotone{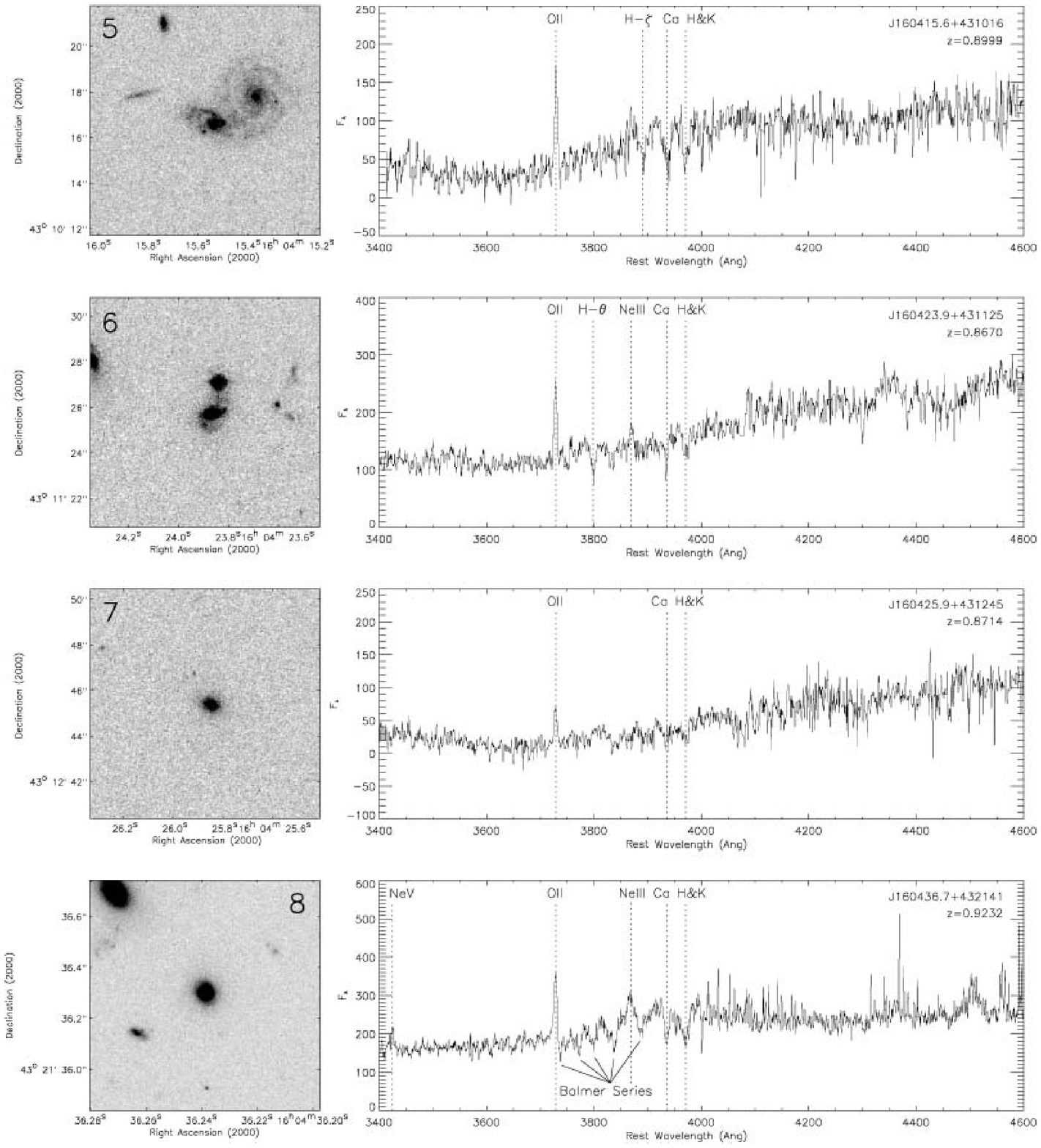}
\caption{(\emph{left}) Thumbnail images of the eight supercluster AGN imaged
         with \emph{HST}/ACS in the F814W band as well as one source imaged with
         \emph{Palomar}/LFC in the $i^{\prime}$ band.  Each image spans roughly
         $10^{\prime\prime}$ on a side and is shown with a log scaling.
         (\emph{right}) Optical spectra of all nine supercluster AGN obtained
         with \emph{Keck}/DEIMOS.  The flux scale is presented in
         self consistent, arbitrary units. The spectra have typically been
         binned by 10 pixels for clarity, where one pixel equals 0.33 \AA.
         \label{fig-thumbs+spect2}}  
\end{figure}

To gain insight into the possible triggering mechanisms of the X-ray-detected
AGN found in the Cl1604 supercluster we examined the morphologies of the host
galaxies within which they reside using both visual and automated techniques.
We limited this analysis to the 8 AGN which are covered by our high-resolution ACS imaging.  Thumbnail
images in the F814W band of each supercluster AGN are show in Figures
\ref{fig-thumbs+spect1} and \ref{fig-thumbs+spect2}.  

Classification of the host morphologies was carried out via visual inspection
by one of the authors (L.M.L.).  The resulting classifications are listed in
Table \ref{tab-prop}.  Also noted are the interpretations of any observed
disturbances using the classification scheme of Smail et al.~(1997).  We find
that a large majority (7/8) of the host galaxies are spheroidal systems.  Five
of the supercluster AGN reside in early-type galaxies (E/S0), two hosts are
bulge dominated systems with faint spiral arms (Sa) and only a single host
is a clear disk dominated, late-type galaxy (Sb). 

These visual classifications were confirmed by the calculation of the Gini
coefficient, $G$, which is a measure of the symmetry in a galaxy's flux
distribution (Abraham et al.~2003) and the $M_{20}$ parameter, which is the
second-order moment of the brightest 20\% of a galaxy's flux and a measure of
central concentration (Lotz et al.~2004).  We measured the parameters using
software written by one of the authors (R.L.) and described in Lin (2007).
We find that although our absolute scaling of both $G$ and $M_{20}$ differ
from that of other software used in the literature (due to a different
implementation and measurement aperture, see Lisker 2008), general trends remain,
with spheroids at lower $M_{20}$ values than disk dominated systems and both
being well separated from merger systems which have higher $G$ values for a given $M_{20}$ range. 

\begin{figure}[t]
\epsscale{1.05}
\plotone{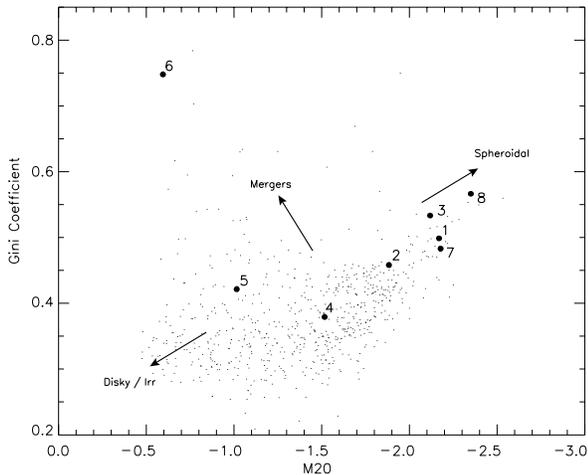}
\caption{Gini and $M_{20}$ parameters for the eight supercluster AGN imaged
  with \emph{HST/ACS} along with a subsample of normal galaxies in the same
  field.  We find the AGN hosts are predominantly early-type or bulge
  dominated galaxies. \label{fig-gini}}
\end{figure}

The location of the 8 host galaxies in the $G-M_{20}$
parameter space is shown in Figure \ref{fig-gini}.  We again find that the supercluster
AGN largely favor early-type hosts.  We find that 5 of 8
host galaxies have parameters consistent with spheroidal and bulge dominated
systems (E/S0/Sa), one source (\#4) lies in the intermediate region occupied by
Sb/bc galaxies, consistent with its visual classification as a spiral galaxy,
and two sources fall within the merger regime.  The two merger
galaxies are sources 5 and 6, which have bright nearby companions.  While it
remains uncertain whether the companion to source 6 is simply seen in
projection, spectroscopy has confirmed that source 5 and its spiral companion 
are indeed a kinematic pair.

Although only two galaxies fall within the merger regime of the $G-M_{20}$
plane, we find indications of merger activity and/or recent interactions in many of
the hosts.  This is most evident in source 2, which appears to be a merger
remnant with a very compact core and faint tidal features.  We suspect the flux contribution from the
tidal features was insufficent to alter the galaxy's symmetry and
concentration enough to move it into the merger regime of the $G-M_{20}$ parameter space.  
Nonetheless, the multiple faint (and blue) tidal
features surrounding the source suggests the system underwent a disruptive
merger event in the recent past.  Faint tidal features are also seen around
source 6 which has a nearby point-like companion.  Furthermore, we find that
source 0, which falls outside of our ACS mosaic, also shows evidence of a
nearby companion in our ground-based imaging.  An image of this galaxy taken
with the Large Format Camera (LFC) on \emph{Palomar} in the Sloan Digital Sky
Survey $i^{\prime}$ filter is shown in Figure \ref{fig-thumbs+spect1};
details of our ground-based imaging and its reduction are described in Gal et al.~(2005).

Other merger indications include double and/or elongated cores.  Source 3
has a partially resolved double core separated by $0.13^{\prime\prime}$ in
projection or $\sim1.0$ $h_{70}^{-1}$ kpc at $z=0.937$.  This separation is similar to the
$z=0.709$ dual-AGN reported by Gerke et al.~(2007) which has a separation of roughly
1.2 $h_{70}^{-1}$ kpc, although here we do not resolve spectral features from
multiple active nuclei.  Despite the presence of multiple cores, the outer
portions of the galaxy appear fairly relaxed with no signs of tidal features
or significant disruptions.  A two-dimension surface brightness fit using the GALFIT software (Peng et
al. 2002) found the galaxy's outer profile to be well fit by a Sersic index
(Sersic 1968) of $n=3.1$, close to a standard de Vaucouleurs profile (de
Vaucouleurs 1948).  This suggests that if the galaxy is the result of a major
merger, sufficient time has elapsed since the event to render the galaxy
largely relaxed.  Source 4 also shows signs of an elongated core, although here we do not
resolve multiple components.

In summary, our visual morphological classifications and those indicated by
the $G$ and $M_{20}$ parameters suggest that the AGN in Cl1604 supercluster
largely favor early-type and bulge dominated host galaxies.  We also find a
majority (6/9) show signs of recent or pending interactions, with one AGN in
a clear merger remnant, three with nearby companions and an additional two with
disrupted cores, despite the fact that only two of these are identified
bona fide mergers by their $G$ and $M_{20}$ parameters alone.  Our results are in good
agreement with the well established finding that AGN are predominantly hosted
by early-type galaxies (e.g.~Kauffman et al.~2003), although we do observe a
larger fraction of disturbed morphologies than was reported by Grogin et
al.~(2005) and Pierce et al.~(2007), who found little evidence for recent
mergers in the hosts high-redshift AGN.  We discuss this disparity in greater depth in  \S\ref{sect-discuss}.

\subsection{Colors}

\begin{figure}[t]
\epsscale{1.1}
\plotone{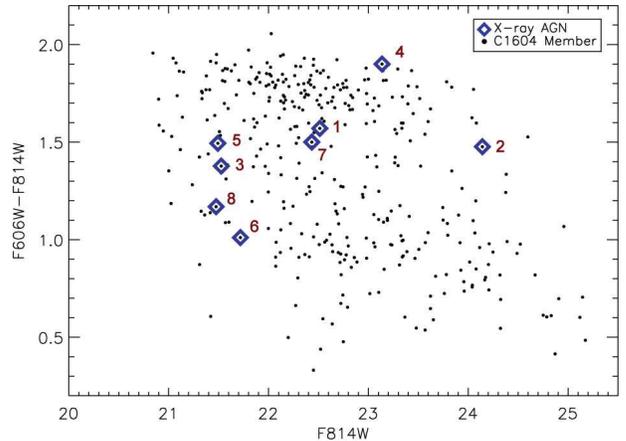}
\caption{Optical color (F606W-F814W) as a function of apparent magnitude in
         the F814W band for the eight supercluster AGN imaged with
         \emph{HST/ACS}.  Also shown are the colors of 363 normal supercluster members.\label{fig-cmd}}
\end{figure}

The optical colors and magnitudes of the 371 supercluster members observed with
ACS in the F606W and F814W bands are show in Figure \ref{fig-cmd}.  The color
bimodality of the galaxies is clearly apparent, with members largely
separated into the supercluster red sequence and a more diffuse blue cloud.  The presence of
such a large number of supercluster members in the blue cloud is a classic
example of the Butcher-Oemler effect (Butcher \& Oemler 1984), where
higher redshift clusters are observed to have an increased fraction of blue,
star forming galaxies relative to the passive cluster populations observed at low redshift.

Also highlighted in Figure \ref{fig-cmd} are the colors of the eight AGN hosts
imaged with ACS. We find that with the exception of source 4, all of the host
galaxies are bluer than the red sequence.  That said, only two sources are
clearly in the blue cloud. Depending on where one defines the boundary of the
blue cloud, we find that 4-5 of 8 hosts have intermediate colors between the
red and blue populations and are located in the transition zone of the
color-magnitude diagram.  Regarding the reddest host, source 4, the galaxy is redder than the
supercluster red-sequence and has the hardest X-ray emission of the 9 AGN,
with $HR = 0.52$. This is indicative of heavy obscuration, suggesting the galaxy's
color is most likely due to heavy extinction rather than a passive stellar
population.  The high fraction of hosts we find in the transition zone
of the color-magnitude diagram is in excellent agreement with the results of
Silverman et al.~(2008), who recently reported an increased fraction of hosts with
intermediate colors in a narrow redshift window containing two large-scale
structures in the E-CDF-S.

Also apparent in Figure \ref{fig-cmd} is that 6 of 8 AGN ($\sim75\%$) are
located in luminous hosts with $m_{814} < 23$.  This translates to a rest
frame $M_{V} = -21.1$ at $z=0.9$ using a $k$-correction appropriate for an Sb
galaxy spectrum.  The preference for AGN to be located in optically luminous
host galaxies has been previously reported in low-redshift superclusters
(Gilmour et al.~2007) and in the field over a wide range of redshifts (Barger
et al.~2003; Nandra et al.~2007; Silverman et al.~2008). Silverman et
al.~(2008) found that $80\%$ of the moderate-luminosity AGN in their sample
reside in galaxies with $M_{V} < -20.7$, in excellent agreement with our
results, despite our small number statistics.

\begin{figure}[t]
\epsscale{1.15}
\plotone{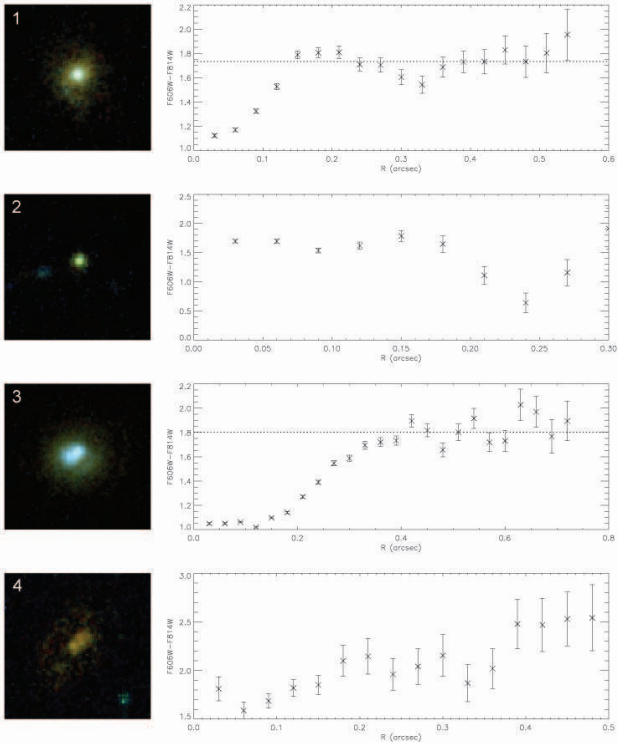}
\caption{(\emph{left}) Color thumbnail images of the eight supercluster AGN
         imaged with \emph{HST/ACS}.  The color components are F606W and F814W for blue and red, respectively, while
         green is represented by a weighted average of the two. (\emph{right}) Spatially resolved color profiles
         (F606W-F814W) of each AGN host.  The profiles are centered on the
         galaxy nucleus and represents the median galaxy color in annuli one
         pixel wide. \label{fig-color-grad1}}
\end{figure}

\begin{figure}[t]
\epsscale{1.15}
\plotone{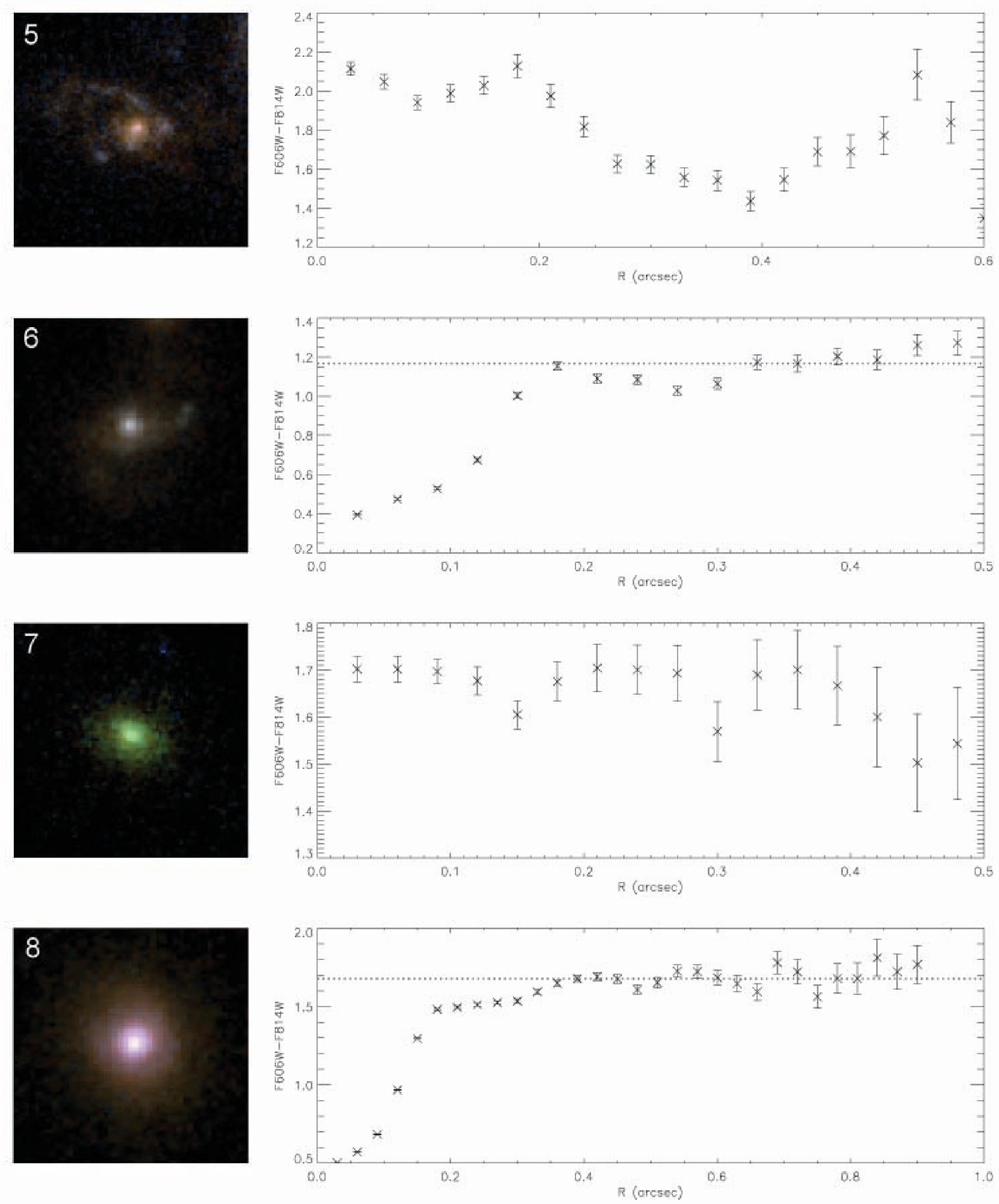}
\caption{(\emph{left}) Color thumbnail images of the eight supercluster AGN
         imaged with \emph{HST/ACS}.  The color components are F606W and F814W for blue and red, respectively, while
         green is represented by a weighted average of the two.
         F606W (blue)...] (\emph{right}) Spatially resolved color profiles
         (F606W-F814W) of each AGN host.  The profiles are centered on the
         galaxy nucleus and represents the median galaxy color in annuli one
         pixel wide. \label{fig-color-grad2}}
\end{figure}

Our high-resolution, dual-band ACS observations also allow us to resolve color
gradients in the AGN host galaxies in order to investigate any recent star
formation activity.  These gradients are shown in Figures \ref{fig-color-grad1} and
\ref{fig-color-grad2}, along with color thumbnails of each host constructed
from observations in the F814W and F606W bands.  We find that four sources
exhibit blue cores similar to those reported by Menanteau et al.~(2001a) and
Treu et al.~(2005).  These include source 1, the double-cored source 3,
source 6 which has a nearby companion and the fairly undisturbed source 8.  

These blue cores may be due to contamination by scattered AGN emission, although we
find this unlikely as i.) the X-ray luminosity of our sources does not reach
the QSO level $(L_{\rm X} \sim 10^{44}$ erg s$^{-1})$ and ii.) the spectra of these four sources show
 no signs of a rising blue continuum (see \S\ref{sect-spect}) indicative of QSO contamination.  
The blue cores are more likely the result of nuclear and/or circumnuclear star formation 
predicted from merger simulations (Mihos \& Hernquist 1996) and observed in low redshift AGN (e.g.~Gu
et al.~2001).  In fact Menanteau et al.~(2001b) find that a merger induced episode of star
formation responsible for producing as little as $20\%$ of the total stellar mass of the system can
lead to blue cores in otherwise red spheroids similar to the galaxies
found by Menanteau et al.~(2001a) and Treu et al.~(2005).  The formation
mechanism proposed by Menanteau et al.~(2001b) is consistent with our
observations as three of the sources have fairly red (F606W-F814W $\sim
1.7-1.8$) colors outside their centers.  We discuss the implications of
this observation further in \S\ref{sect-discuss}.

\subsection{Spectral Properties}
\label{sect-spect}

Spectra of all nine supercluster AGN hosts obtained with the DEIMOS spectrograph are
shown in Figures \ref{fig-thumbs+spect1} and \ref{fig-thumbs+spect2}.  We
find that all of the hosts exhibit narrow [OII] emission lines of low to
moderate strength\footnote{Despite the low signal-to-noise of the source 2
  spectrum and a problematic background subtraction, we do observe a
  double-peaked emission line indicative of [OII] at $z=0.9134$ in the
  extracted spectrum of this host.} .  The equivalent widths (EW) of these
lines, which were measured using the bandpass definitions of Fisher et
al.~(1998), are listed in Table \ref{tab-spect}; the values range from -2.0
to -16.8 \AA.  We also observe that a majority (6 of 8) show Balmer absorption features indicative of young
A-type stars and the lack of massive O and B-type stars.  Two hosts (\#3 and
\#4) show H$\delta$ absorption lines with EWs of 5.2 and 6.0 \AA,
respectively, while a second pair (\#0 and \#8) exhibit a strong higher-order
Balmer series. The most dramatic sets of Balmer lines are seen in sources 3 and
8, two of the blue-cored, early-type hosts.  We note that none of the sources
show signs of broad line emission or significantly rising blue continua commonly found in
more luminous QSOs, suggesting little AGN contamination at shorter wavelengths.

If not for the observed [OII] emission lines, many of the host galaxies would
be classified as post-starburst due to their Balmer absorption. Balmer lines
typically persist for up to $\sim1.0$ Gyr after the last major episode of
star formation due to the lifetime of A-type stars (Poggianti \& Barbaro
1997) and are usually not visible in systems with high levels of ongoing star
formation due to emission filling from HII regions.
Unfortunately given our spectral window we can not use other diagnostic lines
such as H$\alpha$ to determine whether the [OII] features are due to ongoing
star formation or photoionization by the AGN.  If we assume the [OII]
emission is due entirely to star formation, then converting the
[OII] line luminosities to star formation rates (SFR) using the relationship
of Kennicutt (1998) we obtain rates ranging from $0.6\pm0.5$ to $8.5\pm3.9$
M$_{\odot}$ yr$^{-1}$.  All of the derived rates are listed in Table
\ref{tab-spect} and represent lower limits as they are not corrected for extinction.  

While it is more than feasible that many of these galaxies have some ongoing,
centrally-concentrated star formation given their bluer colors, our calculated rates
are likely overestimated (barring extinction corrections) as it is distinctly possible that AGN related
emission contributes to the [OII] line luminosity.  This was recently
demonstrated by Yan et al.~(2006), who found that post-starburst galaxies in
the SDSS sample which show little (or no) H$\alpha$ emission, but have line
ratios consistent with Seyferts, often exhibit significant [OII] EWs. We
have recently undertaken near-infrared spectroscopy of a small subsample of
Cl1604 members using NIRSPEC on \emph{Keck} to determine H$\alpha$ line strengths in
order to perform a similar test at $z=0.9$.  Of the X-ray sources discussed in
this study, only source 5 has thus far been observed.  This galaxy has the highest
measured [OII] EW of the nine hosts and we find an [NII]/H$\alpha$ line ratio
consistent with the system being a transition object, which suggests
both star formation and AGN emission contribute to the system's [OII] line
flux.  Therefore the [OII]-derived SFR for this galaxy is likely an overestimate
as it does not account for the AGN contribution.  We plan to observe the
remaining eight sources, including the four post-starburst hosts, with NIRSPEC in the near future.

It is worth noting that the highest measured SFRs, which are $7.7\pm3.5$ and
$8.5\pm3.9$ M$_{\odot}$ yr$^{-1}$, are found in sources 3 and 8,
respectively; the two galaxies with the most pronounced Balmer absorption.
It is unlikely that these absorption lines would be so prevalent in systems
that are forming stars as prodigiously as these rates suggest.  Poggianti et
al. (1999) found that spectra with strong Balmer lines and measurable [OII]
can be reproduced with low-level star formation following a recent, strong
burst.  This may in fact be what we are observing here if our [OII]-derived SFRs
are overestimated due to AGN contaimination. Alternatively the [OII] emission
may originate entirely from gas ionized by the AGN. In either case it would
seem a large fraction of our supercluster AGN are located in galaxies which
underwent a starburst phase within the last $\sim1$ Gyr and whose star
formation has since been truncated.  

In addition to the post-starburst features, most hosts also show Ca H\&K
lines and three sources have a clear continuum break at 4000\AA, features indicative of an underlying older stellar
population.  This is observed in the spectra of sources 3, 6 and 8, all
of which are blue-cored galaxies with otherwise red colors.  This set of
observed properties is consistent with a minor merger induced nuclear
starburst in a previously passive system, similar to the process described by
Menanteau et al.~(2001b).

\subsection{Environments}
\label{sect-enviro}

The location of the nine AGN detected in the Cl1604 supercluster in relation
to the system's constituent clusters are shown in Figure
\ref{fig-agn-spatial-dist}.  We find the greatest concentration of AGN near
Cl1604+4316 (hereafter Cluster C) at $z=0.935$.  Four sources have redshifts
nearly equal to the cluster's systemic redshift (\#0, \#1, \#3, \#4), while a
single source is located in the foreground of the cluster (\#2 at $z=0.913$).
Three of the AGN are located on the outskirts of Cl1604+4314 (Cluster B) at
$z=0.865$; the two sources closest to the cluster center have nearly
identical redshifts to that of the cluster (\#6 and \#7), while the one with
the greatest projected distance is at a higher redshift (\#5 at $z=0.899$).
In Cl1604+4321 (Cluster D) we find a single AGN (\#8) with a redshift
identical to that of the systemic cluster redshift of $z=0.923$.  
 
\begin{figure}[t]
\epsscale{1.1}
\plotone{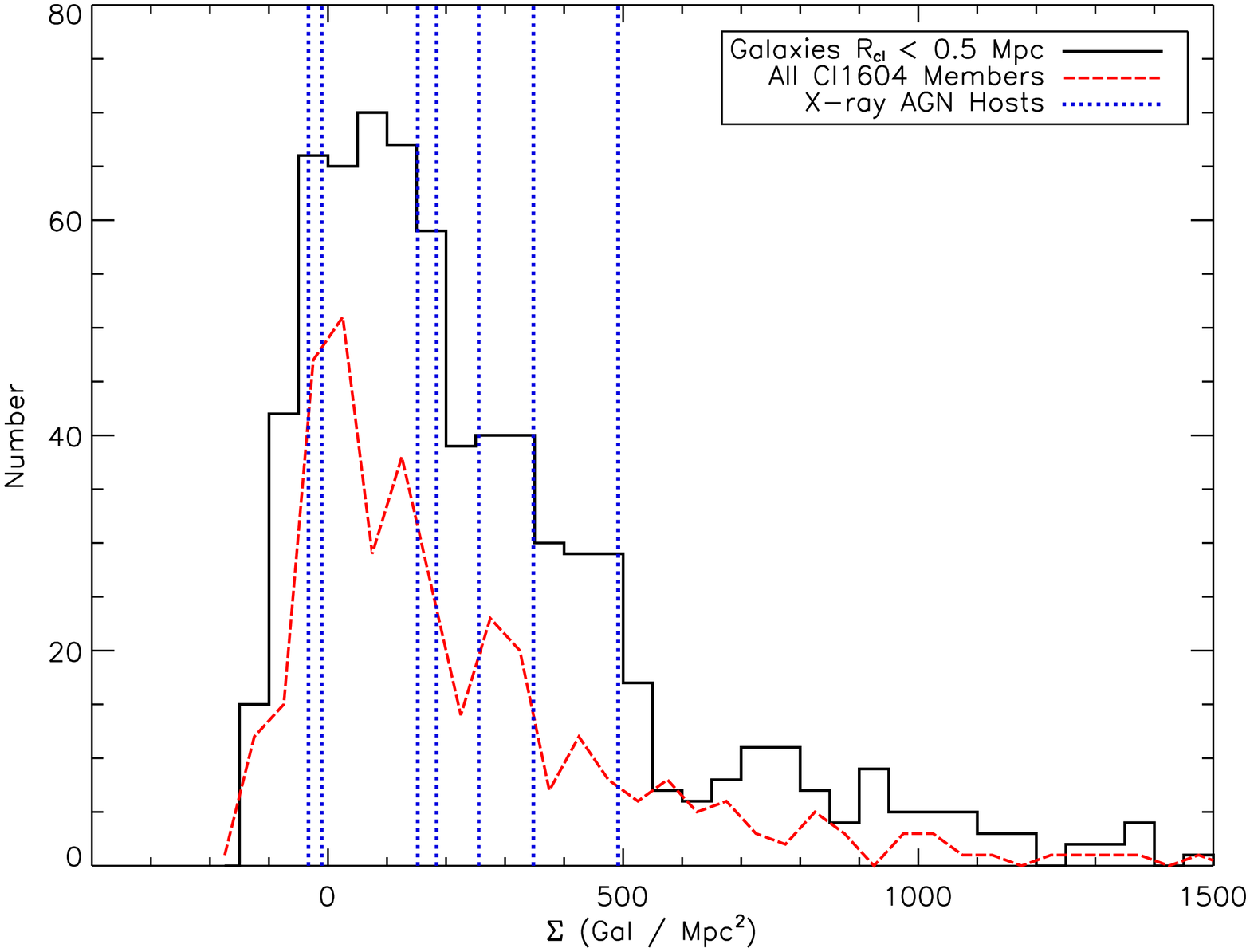}
\caption{Projected galaxy density above the field density in the
  direction of the Cl1604 Supercluster.  Densities were calculated using all
  galaxies down to $M_{V} = -19.27$ ($m_{814} = 24.82$).  The density
  distribution for supercluster members and X-ray AGN hosts are denoted by
  the dashed red and vertical dotted magenta lines, respectively.  The solid
  histogram shows the density of all ACS detected galaxies within 0.5 $h_{70}^{-1}$ Mpc of
  a cluster center.  \label{fig-den7}}
\end{figure}

A striking result is the general absence
of AGN in regions with the highest local galaxy densities.   This includes
Cl1604+4304 (Cluster A) and the center of Cluster B.  These systems are
the richest clusters in the Cl1604 complex, with the highest measured
velocity dispersions and X-ray luminosities (G08; K08).  On the other hand,
we do find AGN near the centers of Cluster C
and D (both in projection and redshift space), both of which are poorer
systems that are not directly detected in our X-ray observations.   

The distribution of AGN in the supercluster suggests their activity may be
influenced by both local environment and more large-scale cluster properties
such as a system's dynamical state.  This is highlighted by the presence of
AGN on the outskirts of Cluster B and the lack of a similar population around
Cluster A.   While both systems are relatively massive, G08
found evidence of velocity segregation in the redshift distribution of
Cluster B, indicating the system may have undergone a recent merger or is
actively accreting a substantial number of galaxies.  K08
also found the system to have a higher than expected velocity dispersion
relative to the $L_{\rm x}-\sigma_{v}$ relation, also suggesting the system
may not be fully relaxed.  This is quite unlike Cluster A, which shows no
signs of recent merger activity and appears fairly relaxed (and consistent
with the $L_{\rm x}-\sigma_{v}$ relation).   We speculate
that the difference in the dynamical activity of Clusters A and B may indicate
the global properties of these systems are affecting the level of AGN
activity observed around each cluster. 

In order to quanitify the local environments of the AGN and to compare
them to those of the entire supercluster population we calculated the local
projected galaxy density, $\Sigma$, near each host galaxy using the method employed by Dressler et
al.~(1980) and more recently by Smith et al.~(2005) and Postman et
al.~(2005).  The method calculates the projected density as $\Sigma = N / \Omega_{N}$,
where $\Omega_{N}$ is the solid angle which encompasses a galaxy's $N$
nearest neighbors down to a certain absolute magnitude limit. For our
calculation we adopt the parameters used by Postman et al.~(2005) with
$N=7$ and an absolute $V$-band magnitude of $M_{V} = -19.27$.  At $z=0.9$, the average
redshift of the supercluster, and using an Sb galaxy $k$-correction, the latter
translates to an observed magnitude of $m_{814} = 24.82$.  We found that varying
these parameters in the range of $5 \le N \le 10$ and $24 \le m_{814}
\le 25$ did not significantly alter our results.

Since the Cl1604 field lacks sufficient redshift information to calculate
$\Sigma$ using only confirmed supercluster members, we make use of all
galaxies falling above our magnitude limit to determine the projected
density and statistically correct for background and foreground
contamination.  This correction is derived from the average surface density of
field galaxies with $m_{814} \le 24.82$ in seven blank fields observed with
ACS.  These fields were chosen for their lack of cataloged structures and
include three fields from GO-9761 (PI Alamini) and four fields from GO-9468
(PI Yan)\footnote{The field centers are 03:42:48.37 -09:36:51.9, 09:54:34.40
  +17:53:11.2, 10:01:46.32 +55:05:24.7, 16:36:58.01 +40:56:01.1, 16:37:04.11
  +40:58:59.1, 16:37:04.11 +41:01:58.1, 21:55:42.97 -09:26:34.7.}.   Each field was obtained from the HST archive and reduced in an
identical manner to that of the Cl1604 dataset.   Since we correct for the background and foreground
contamination by subtracting the average surface density of a blank field our
projected densities differ from those of Smith et al.~(2005) and Postman et
al.~(2005) in that they are technically densities in excess to the field
densities.  In other words, our projected densities have been normalized to
the field density at $z=0.9$.

The distribution of projected densities for all 371 supercluster members
imaged by ACS is shown in Figure \ref{fig-den7}.  Also shown are the local
densities of the eight AGN host galaxies.  We find two sources (\#4 and \#5)
have densities similar to that of the field, which given our statistical correction is
set to $\Sigma = 0$ galaxies Mpc$^{-2}$.  The remaining sources are located in regions of
intermediate density, with $\Sigma$ values larger than the field value, but well below the
densities found near the cluster centers which range from 700-800 galaxies
Mpc$^{-2}$ in the center of cluster C and D and $\sim1000$ galaxies Mpc$^{-2}$
in Cluster A and B.  These densities confirm our conclusions determined by
visual inspection: the AGN are found in intermediate density environments,
either outside the centers of massive clusters or in poorer clusters.  It has
long been thought that such regions are conducive to galaxy interactions due
to their higher densities compared to the field and their relatively low
velocity dispersions (Cavaliere et al.~1992).  This finding is similar to the conclusions of Gilmour et al.~(2007) who found
AGN in the Abell 901/902 supercluster at $z\sim0.17$ avoid the densest environments
in the system and are instead located on the outskirts of clusters or in
galaxy groups.  At higher redshifts, we also agree with the results of
Georgakakis et al.~(2007) who found AGN generally avoid underdense regions and are found in
richer environments than similar non-active galaxies.

\section{Cluster AGN Fraction}

There are several lines of evidence that suggest AGN activity is triggered more often in
and around galaxy clusters than in the field.  A growing number of observations have reported an
overdensity of X-ray point sources in the vicinity of clusters relative to
blank-field observations (Henry \& Briel 1991; Cappi et al.~2001; Ruderman \& Ebeling
2005; K08) and there are indications this overdensity
increases with redshift (Cappelluti et al.~2005; Branchesi et al.~2007).
This trend was recently confirmed spectroscopically by Eastman et al.~(2007),
who found an order-of-magnitude increase in the number of cluster members hosting AGN in
four $z\sim0.6$ clusters compared to the fraction measured by Martini et
al.~(2006) in eight low-redshift clusters ($z < 0.31$) using the same
methodology.  We can perform a similar analysis using the AGN in the Cl1604 supercluster to
investigate whether this trend continues out to $z\sim0.9$.

 Eastman et al.~(2007) found a total of eight galaxies hosting AGN with X-ray luminosities of
 $L_{\rm X} > 10^{43}$ $h_{70}^{-2}$ erg s$^{-1}$ (2-10 keV) within the $R_{200}$ radius of four
 higher-redshift clusters.  This was out of $\sim400$ potential hosts brighter
 than $M_{R} = -20$, resulting in an AGN fraction of $f_{A}(L_{\rm X} >
 10^{43};M_{R} < -20) = 0.020_{-0.008}^{+0.012}$.  To calculate the $R_{200}$ radii of the
 three clusters in the Cl1604 system which have associated AGN, we make use of the
 velocity dispersions, $\sigma_{v}$, measured by G08 and the
 equation $R_{200} = 2\sigma_{v} / [\sqrt{200} H(z)]$.  The resulting
 $R_{200}$ radii range from 0.52 to 1.01  $h_{70}^{-1}$ Mpc (or $1\farcm11$ to
 $2\farcm18$ on the sky).  Within these radii we find three supercluster AGN,
 although only two have 2-10 keV X-ray luminosities greater than $10^{43}$
 $h_{70}^{-2}$ erg s$^{-1}$.  These are sources 7 and 8 in clusters B and D.

To determine which non-active cluster galaxies are capable of hosting a
detectable AGN we construct a flux limit map covering our entire ACIS-I
mosaic.  This map takes into account the local background level, the size of the
PSF and the detector vignetting at the location of each cluster member to
determine the faintest detectable AGN at that position; details on its
construction are given in K08. Within the $R_{200}$ radii
of clusters B and D we find 74 galaxies brighter than $M_{R} = -20$ within
which both AGN could have been detected.  Rest-frame magnitudes for the
cluster members were calculated from our observed magnitudes in the F814W
band using an Sb galaxy $k$-correction\footnote{We also
  convert our magnitudes from the AB to the Vega system using the
  appropriate ACS zeropoint in order to match the magnitude limit of Eastman et al.~(2007).}.  

The resulting AGN fraction amounts to 2 AGN in 74 possible hosts or $f_{A}(L_{\rm X} >
10^{43};M_{R} < -20) = 0.027_{-0.020}^{+0.045}$ (the uncertainties correspond to
90\%, one-sided Poisson confidence limits). An important point to note is that we have treated the diffuse emission from
Cluster B as an enhanced local background in constructing our flux limit map.
Although this emission effectively raises our point source detection
threshold in the soft band, the emission does not affect our ability to detect sources in
the hard band.  We calculate that both AGN detected within the
$R_{200}$ radius of clusters B and D would have been detected in all of the
cluster members with $M_{R} < -20$ out to the same radius.   If we instead
wish to be conservative and remove all galaxies embedded in the cluster's
diffuse emission, which extends roughly $50^{\prime\prime}$, this reduces the
number of potential hosts by 14, resulting in an AGN fraction of $f_{A}(L_{\rm X} >
10^{43};M_{R} < -20;$ No Cluster B$) = 0.033_{-0.025}^{+0.055}$.  

Our measured cluster AGN fraction within $R_{200}$ at $z\sim0.9$ is
significantly higher than the fraction measured at low-redshift by Martini et
al.~(2006), who found $f_{A}(L_{\rm X} > 10^{43};M_{R} < -20) =
0.0007_{-0.0007}^{+0.0021}$, but largley consistent with the moderate-redshift results of
Eastman et al.~(2007).  Even if we exclude galaxies embedded within the
diffuse emission of cluster B, we do not find a statistically significant
increase over the fraction measured at $z\sim0.6$ given our large errors ($0.033_{-0.025}^{+0.055}$
versus $0.020_{-0.008}^{+0.012}$).  Therefore we can not report a substantial
change in AGN activity within the $R_{200}$ radius from $z\sim0.6$ to
$z\sim0.9$.  This is perhaps not surprising
since, as we discussed in \S\ref{sect-enviro}, most of the AGN detected in
the supercluster (6 of 9) are found in intermediate density environments and
well outside the $R_{200}$ radius of its constituent clusters. A much larger
sample of AGN in $z\sim1$ clusters will be needed in order to determine
whether the AGN fraction increases substantially beyond the levels observed
at moderate redshifts.

\section{Discussion}
\label{sect-discuss}

In many cases the observed properties of the nine AGN hosts found in
the Cl1604 supercluster either confirm the findings of previous wide-field
galaxy surveys or extend to higher redshifts trends observed in
lower-redshift structures.  With regard to overall host morphology, optical
luminosity and color, our findings are generally consistent with the results
of several past studies which find host galaxies to be largely massive,
early-type galaxies showing signs of recent star formation (Kauffman et
al.~2003; Heckman et al.~2004, Sanchez et al.~2004; Nandra et al.~2007;
Silverman et al.~2008).  The fact that our supercluster AGN are found in
luminous spheroids which have bluer colors than similar non-active galaxies
is not surprising as it is precisely these galaxies that contain the two
ingrediants required for nuclear activity: a massive black hole and an
abundant gas supply (Kauffmann et al.~2003).  Our finding of predominantly
blue-cored, early-type hosts agrees well with the growing consensus that
powerful AGN are located in massive galaxies with bulges containing younger
stellar populations.

Our results also seem to confirm the increased post-starburst-AGN
connection reported at higher redshifts.  While there is extensive evidence for
younger stellar ages and recent starburst activity in moderate-luminosity AGN
(Schmitt et al.~1999, Gonzalez Delgado et al.~2001; Cid Fernandes et
al.~2001; Kauffmann et al.~2003; Heckman 2004; Ho 2005), the fraction of
Seyferts associated with post-starburst hosts at low redshift remains fairly
small. Goto et al.~(2006) found that although post-starburst hosts were more
often associated with AGN than normal galaxies, hosts exhibiting strong
post-starburst spectral signatures were relatively rare, accounting for only
4.2\% of narrow-line type 2 Seyferts in the SDSS sample.  Recently
Georgakakis et al.~(2008) reported an increase in this fraction at higher
redshifts ($0.68 \le z \le 0.88$) using a sample of X-ray sources detected in
the Extended Groth Strip.  They find that $21\%$ of AGN hosts located on the red sequence
exhibit post-starburst signatures.  In the Cl1604 supercluster we find
a significant fraction ($\sim44\%$) of all host spectra show post-starburst
features, supporting the notion of an increased post-starburst-AGN association
at redshifts approaching unity.  Of course we must use caution when directly
comparing these results as we have assumed at least some AGN emission
contributes to our observed [OII] lines whereas previous studies have
not.  In fact Yan et al.~(2006) warn that AGN driven [OII] emission being
mistaken for ongoing star formation may have led previous studies to
underestimate the prevalence of the post-starburst-AGN connection at lower redshifts.
 
On one important topic, merger activity as a triggering mechanism, our
observations do seem to be at odds with previous studies.  Although luminous QSO
activity has been tied to merger activity (Canalizo \& Stockton 2001,
Bennert et al.~2008), a similar link has not been clearly established with
moderate-luminosity Seyferts (De Robertis et al.~1998; Grogin et al.~2003, 2005; Pierce et
al.~2007).  In particular Grogin et al.~(2005) found that while higher-redshift hosts are
bulge dominated, their asymmetry and near-neighbor counts were consistent
with non-AGN control samples.  We find that although only two of eight hosts
have asymmetry and concentration parameters indicative of ongoing mergers, several
galaxies do show signs of recent interactions (i.e.~tidal features or
assymetric/multiple cores).  Many of these hosts appear otherwise relaxed (being fit with high Sersic
indicies, for example), implying that any merger activity was either fairly
minor so as to not disrupt the galaxy significantly or enough time has
transpired for the system to have sufficiently relaxed.  That being said, we
note once again that the asymmetry of the single AGN that we find in an apparent
merger remnant was not greatly enhanced by the precence of faint tidal
features largely due to its considerably brighter and compact nucleus.  

Overall we find that the properties of the nine host galaxies are generally
consistent with a scenario in which recent interactions have triggered both
increased levels of nuclear activity and an enhancement of centrally
concentrated star formation, followed by a rapid truncation of the latter.   
This is evidenced by the post-starburst signatures in several host spectra,
their blue-cored color morphologies, and the high fraction of hosts that show signs of
recent interactions.  Whether the star formation and nuclear activity were
triggered simultaneously or with some delay remains to be determined, but the
AGN emission does seem to persists beyond the termination of starburst activity in these systems. 
Given this fact, our observations are therefore consistent with models
in which feedback from the AGN itself is responsible for this supression
and the subsequent migration of these galaxies from the blue cloud to the red
sequence.  We do find that a majority of the host galaxies have intermediate
colors that place them in the transition zone of the color-magnitude
diagram, which further supports this scenario.   This is in agreement with
several recent studies which have also found that the color of host galaxies
cover a similar range, extending from the red edge of the blue cloud to the red sequence (Sanchez
et al.~2004; Nandra et al.~2007; Martin et al.~2007; Georgakakis et al.~2008;
Silverman et al.~2008; although see Westoby et al.~2007 for a differing conclusion).

Of particular interest for our study are the results of Silverman et
al.~(2008), who found that within a redshift interval containing two
large-scale structures in the E-DF-S $(0.63 \le z \le 0.76)$ AGN hosts are
preferentially located in the transition zone.  We find a similar result
within the Cl1604 complex, where a majority of host galaxies are located
between the blue cloud and the red sequence.  These results suggest that the
hosts of moderate-luminosity AGN are predominantly a transition population
within these structures, caught in the process of migrating from blue to red.
Coupled with recent findings that AGN are more common in richer environments
than they are in the field (Gilli et al~2003; Cappelluti et al~2005; Eastman
et al.~2007; K08), this finding implies that AGN related
feedback may play an important role in accelerating galaxy evolution in
large-scale structures.

That being said, we should note that in studying the low-redshift supercluster Abell 901/902,
Gilmour et al.~(2007) found many AGN hosts with colors similar to that of the
supercluster red sequence. Likewise Nandra et al.~(2007) and Silverman et
al.~(2008) found a significant population of field AGN associated with redder
hosts.  As Silverman et al.~points out, most of these are found at lower-redshifts
where the strong color evolution of massive galaxies capable of hosting AGN
insures these galaxies will be redder than those found at $z=0.9$.  We would
like to further point out that most of these galaxies harbor
AGN less luminous than those detected in the Cl1604 supercluster.  For
example, a source detected at our $3\sigma$ flux limit would have a
rest-frame 0.5-8 keV luminosity of $\sim8\times10^{42}$ $h_{70}^{-2}$ erg
s$^{-1}$ at the average redshift of the Cl1604 supercluster.  All of the X-ray sources detected by Gilmour et al.~(2007) have
luminosities below this limit.  Likewise most of the sources detected by
Nandra et al.~(2007) on the red sequence have luminosities below $10^{43}$
$h_{70}^{-2}$ erg s$^{-1}$; only one of our AGN fall into this regime.  This
is not extremely surprising as it has been observed for some time that less
luminous AGN have older stellar populations than their more powerful
counterparts (e.g. Kauffmann et al.~2003; Heckman 2004).  The fact that we do
not observe any moderate-luminosity AGN on the red sequence is consistent
with the AGN feedback model of Hopkins et al.~(2005) which suggests AGN
should become less luminous as their hosts migrate from the blue cloud to the
red sequence due to the depletion of material which would otherwise feed the
central black hole.

An alternative theory that is also consistent with our observations is one of
episodic star formation and AGN activity in red sequence galaxies.  In this
scenario minor mergers that do not significantly disrupt the galaxy initiate
nuclear activity and trigger renewed star formation which temporarily moves
the system off of the supercluster red sequence. This produces a population
of young, blue stars we observe as the blue cores of these systems in a
process similar to the one described by Menanteau et al.~(2001b).  This
results in a build-up of both stellar bulge mass and black hole mass that is
eventually regulated by the feedback mechanism.  As star formation is
quenched, the newly formed stars fade and the galaxy returns to the red
sequence but this time with a higher luminosity.  In this view, AGN hosts in the
supercluster are not in fact transitioning from blue to red, but instead from
low to high luminosity (mass) on the red sequence.  This scenario does seem to be supported by
the relatively red colors we find in many of the blue-cored host galaxies.
If it were not for their cores, these galaxies would be on the red sequence.
Of course in order for this process to work, either the red sequence galaxies would
need to retain a significant reservoir of cold gas to fuel future activity or
a sufficient amount would need to be accreted from the merging, presumable
gas-rich, galaxy.

\section{Conclusions}

In summary, we find that the X-ray selected AGN detected in the Cl1604
supercluster are largely hosted by luminous ($M_{V} < -21.1$), spheroidal
and/or bulge dominated galaxies which are bluer than 
similar galaxies on the system's red sequence.
The integrated colors of a majority of the hosts (5/8) place them in the
valley between the supercluster red sequence and the more diffuse blue cloud.
In half of the hosts, the bluer overall colors are the result of blue
resolved cores in otherwise red spheroids.  Given that the supercluster AGN
luminosities do not reach the QSO level and the lack of obvious AGN contamination
in their optical spectra, we interpret the blue cores as the result of recent
star formation.  Many of these galaxies do show signs of starburst
activity within the last $\sim1$ Gyr as $\sim44\%$ (4/9) of the host spectra exhibit
either strong H$-\delta$ absorption or a pronounced Balmer series.  We also
find a majority of hosts (6/9) show signs of recent or pending
interactions, a possible indication of their triggering mechanism.  With
regard to environment, we observe that the hosts tend to avoid the densest regions of the
supercluster and are instead located in intermediate density environments,
such as the infall region of a massive cluster or in poorer systems
undergoing assembly.  Finally, we measure an increased fraction of
supercluster members that harbor AGN relative to low-redshift ($z<0.3$) clusters,
but we do not observe a substantial change from the fraction reported at
$z\sim0.6$.

 In general our observations are fairly consistent with theories that link
 AGN activity to galaxies undergoing transformation from star forming systems in the
 blue cloud to passively evolving galaxies on the red sequence.   This is
 evidenced by the fact that (i) a significant fraction of the AGN hosts are
 located in the transition zone of the color-magnitude diagram, which
 is normally sparsely populated, (ii) several hosts show evidence of a recent enhancement of star
 formation that was abruptly terminated and (iii) there are indications of
 recent merger activity in many of the hosts, which are located in
 moderate-density environments thought to be conducive to galaxy interactions.
 An alternative theory that is also consistent with our observations is one of
 episodic star formation and AGN activity in red sequence galaxies triggered
 by galaxy interactions.  In this scenario, AGN hosts in the
 supercluster are not in fact transitioning from blue to red, but instead from
 low to high luminosity (mass) on the red sequence.

Many of these conclusions, especially those regarding the post-starburst-AGN connection, depend on our assumption that the [OII]
emission lines seen in the host galaxies are due in part to AGN emission.  We
plan to test this assumption in the near future with near-infrared spectroscopic
observations of each AGN host in order to obtain H$\alpha$ and [NII] EWs and
line ratios.  The observations of one host which have already been completed have confirmed
that AGN emission contributes to the galaxies [OII] line flux, leading to an
overestimated SFR.  If we find similar results for the remaining hosts, this
will help confirm that a significant fraction of AGN in the Cl1604
supercluster have experienced a recent and rapid truncation of their star
formation activity.

\acknowledgments
This work is supported by the Chandra General Observing Program
under award number GO6-7114X. Additional support for this program was
provided by NASA through a grant HST-GO-11003 from the Space Telescope
Science Institute, which is operated by the Association of Universities for
Research in Astronomy, Inc. The
spectrographic data used herein were obtained at the W.M. Keck Observatory,
which is operated as a scientific partnership among the California Institute
of Technology, the University of California and the National Aeronautics and
Space Administration. The Observatory was made possible by the generous
financial support of the W.M. Keck Foundation.


\begin{center}
\tabletypesize{\scriptsize}
\begin{deluxetable*}{ccccccc}
\tablewidth{0pt}
\tablecaption{X-ray Properties of X-ray Selected AGN in the Cl1604 Supercluster \label{tab-prop}}
\tablecolumns{7}
\tablehead{\colhead{} & \colhead{} & \colhead{RA} & \colhead{Dec} &
           \colhead{} &  \colhead{$L_{\rm x}$ (0.5-8 keV)} & \colhead{} \\  
           \colhead{ID} & \colhead{Name} & \colhead{(J2000)} &  \colhead{(J2000)} &  
           \colhead{$z$} & \colhead{($\times10^{43}$)$^{\dagger}$} & \colhead{HR} } 
\startdata
 0  & J160401.3+431351  &  241.00546  &  43.23089  &  0.927  &  3.44  & -0.31  \\   
 1  & J160405.1+431519  &  241.02139  &  43.25551  &  0.934  &  0.78  & -0.50  \\    
 2  & J160406.1+431807  &  241.02528  &  43.30222  &  0.913  &  2.99  & -0.36  \\   
 3  & J160408.2+431736  &  241.03416  &  43.29359  &  0.937  &  2.75  &  0.12  \\   
 4  & J160410.9+432111  &  241.04547  &  43.35318  &  0.935  &  1.12  &  0.52  \\    
 5  & J160415.6+431016  &  241.06492  &  43.17134  &  0.900  &  4.47  & -0.35  \\   
 6  & J160423.9+431125  &  241.09959  &  43.19052  &  0.867  &  3.39  & -0.42  \\   
 7  & J160425.9+431245  &  241.10783  &  43.21265  &  0.871  &  1.31  & -0.42  \\   
 8  & J160436.7+432141  &  241.15290  &  43.36147  &  0.923  &  2.28  & -0.10  \\   
\vspace*{-0.075in}
\enddata
\tablecomments{$^{\dagger}$ in units of $h_{70}^{-2}$ erg s$^{-1}$}
\end{deluxetable*}
\end{center}

\begin{center}
\tabletypesize{\scriptsize}
\begin{deluxetable*}{cccccccl}
\tablewidth{0pt}
\tablecaption{Optical Properties of X-ray Selected AGN in the Cl1604 Supercluster \label{tab-spect}}
\tablecolumns{8}
\tablehead{\colhead{}   & \colhead{}     & \colhead{[OII] EW} &  \colhead{H$\delta$ EW} & \colhead{SFR} & \colhead{Morph} & \colhead{} & \colhead{} \\  
           \colhead{ID} & \colhead{Name} & \colhead{(\AA)} & \colhead{(\AA)} & \colhead{M$_{\odot}$ yr$^{-1}$}  & \colhead{Class} & \colhead{Int$^{\dagger}$} & \colhead{Notes}}
\startdata
 0  & J160401.3+431351  &  -3.3  & 1.8  &  2.7$\pm$1.3  &  -    & -  & No ACS data available \\   
 1  & J160405.1+431519  &  -2.5  & ---  &  1.6$\pm$0.9  &  E    & -  & Blue core, faint elongation, possible tidal feature \\    
 2  & J160406.1+431807  &  -2.0  & ---  &  0.6$\pm$0.5  &  E/X  & T  & Merger remnant, tidal features \\   
 3  & J160408.2+431736  &  -7.2  & 5.2  &  7.7$\pm$3.5  &  Epec & M  & Blue, double peaked core  \\   
 4  & J160410.9+432111  &  -5.2  & 6.0  &  0.9$\pm$0.6  &  Sa   & -  & Red galaxy with elongated core \\    
 5  & J160415.6+431016  & -16.8  & ---  &  4.5$\pm$2.1  &  Sb   & M  & Spiral with kinematic companion \\   
 6  & J160423.9+431125  &  -7.5  & ---  &  4.3$\pm$2.0  &  Sa   & M  & Blue core, nearby companion, tidal features \\   
 7  & J160425.9+431245  & -13.7  & ---  &  1.3$\pm$0.7  &  E    & -  &  \\  
 8  & J160436.7+432141  &  -7.0  & ---  &  8.5$\pm$3.9  &  E    & -  & Blue core, appears relaxed \\   
\vspace*{-0.075in}
\enddata
\tablecomments{ $^{\dagger}$ Interpretation of disturbance: T = Tidal Feature, M = Merger}
\end{deluxetable*}
\end{center}

\clearpage

\bibliographystyle{apj}
\bibliography{/Users/kocevski/LaTex/ORELSE/SuperclusterII/ms}

\end{document}